\documentclass[12pt,a4paper]{article}
\usepackage[utf8]{inputenc}
\usepackage[T1]{fontenc}
\usepackage[top=2.5cm, bottom=2.5cm, left=2.0cm, right=2.0cm]{geometry}

\usepackage[section]{placeins}		

\usepackage{amscd,amsfonts,amsmath,amssymb,amsthm,bbm,bm,latexsym,mathrsfs,mathtools,thmtools}
\usepackage[subnum]{cases}	
\allowdisplaybreaks			
\makeatletter
\newcommand*{\distas}[1]{\mathbin{\overset{#1}{\kern\z@\sim}}}	
\makeatother

\theoremstyle{definition}
\newtheorem{definition}{Definition}

\theoremstyle{remark}

\newtheorem{example}{Example}

\usepackage{hyperref} 		
\hypersetup{bookmarks=true, unicode=true, urlcolor=blue, linkcolor=blue, filecolor=blue, colorlinks=true, citecolor=black, final=true}

\usepackage{xcolor,subfigure,epsfig,graphics,graphicx}
\usepackage{caption}
\usepackage{enumitem}
\usepackage{longtable,float,array}
\usepackage{setspace}
\usepackage{threeparttable,booktabs,rotating}

\makeatother
\newcolumntype{C}[1]{>{\centering\arraybackslash}p{#1}}

\begin{document}


\title{\textbf{On the ``mementum'' of Meme Stocks}\thanks{\footnotesize 
Matteo Iacopini acknowledges financial support from the EU Horizon 2020 programme under the Marie Skłodowska-Curie scheme (grant agreement no. 887220).}}

\author{
Michele Costola\thanks{Ca' Foscari University of Venice, Italy. \color{blue}\texttt{michele.costola@unive.it}}
\and
Matteo Iacopini\thanks{Vrije Universiteit Amsterdam and Tinbergen Institute, The Netherlands. \color{blue}\texttt{m.iacopini@vu.nl}}
\and
Carlo R.M.A. Santagiustina\thanks{Ca' Foscari University of Venice and Venice International University, Italy. \color{blue}\texttt{carlo.santagiustina@unive.it}}
}

\maketitle





\begin{abstract}
The meme stock phenomenon is yet to be explored. In this note, we provide evidence that these stocks display common stylized facts on the dynamics of price, trading volume, and social media activity. Using a regime-switching cointegration model, we identify the meme stock ``mementum'' which exhibits a different characterization with respect to other stocks with high volumes of activity (persistent and not) on social media.
Understanding these properties helps the investors and market authorities in their decision.

\noindent%
{\bf Keywords:} Meme stocks; Social media; Social trading; Cointegration; Regime switching
\end{abstract}


\section{Introduction}

Recently, the ``meme stock'' phenomenon has drawn major attention as social media have been used as coordination devices to synchronize on buying signals that have significantly affected the price and trading volumes of certain stocks.
Similar to cryptocurrencies, many retail investors have jumped on the meme stock bandwagon even in absence of meaningful changes of the value of the fundamentals.
However, this phenomenon is characterized by non-traditional drivers.
First, the majority of the actors initially involved in the aforementioned process are retail investors.
Second, these investors use social media to voluntarily and openly coordinate their stock purchasing decisions. This contrasts with the fundamentals strategies adopted by institutional market operators such as mutual and hedge funds.

The most famous episode occurred in January 2021 and involved the US video game retailer GameStop in a short squeeze conceived by the users of the \textit{r/wallstreetbets} subreddit in the online social platform Reddit. As a consequence of the above-mentioned coordination mechanism, the trading strategy has caused a significant increase in the stock price which has reached approximately 140\% in one single market day. On the other hand, it has provoked financial losses on several hedge funds that were pursuing a short position strategy due to the worsening of the firm's fundamentals.\footnote{For instance, see \href{https://bloomberg.com/news/articles/2020-12-08/gamestop-declines-after-sales-fall-more-than-analysts-estimated}{https://bloomberg.com/news/articles/2020-12-08/gamestop-declines-after-sales-fall-more-than-analysts-estimated}.} The scheme has last for several days.

The fact that retail investors may be active on several platforms simultaneously and that subreddits, like \textit{r/wallstreetbets}, are directly linked to official profiles in other social media\footnote{For example, see \textit{r/wallstreetbets} on Twitter: \href{https://twitter.com/Official\_WSB}{twitter.com/Official\_WSB}, and Discord:  \href{https://discord.com/invite/wallstreetbets}{discord.com/invite/wallstreetbets}.} makes their activity natively multi-platform, thus implying that buying signals can be ubiquitous.
This reinforces the rippling of meme stock phenomena across social media and their audiences.
For example, in mid-January 2021 some users were posting both buying signals about GameStop on Reddit, via \textit{r/wallstreetbets}, and also on Twitter.
Besides, their openness makes social media ideal instruments for decentralized, multi-platform coordination processes.
Summarizing, the meme stock phenomenon cannot be considered platform-specific and its reach far exceeds \textit{r/wallstreetbets} and Reddit itself.

In this letter, we provide the first evidence that meme stocks share common stylized facts on the dynamics of price, trading volume, and social media activity on Twitter.
We investigate the ``\textit{mementum}'', or meme period, of stocks, that is the momentum when synchronized buying signals originated on social media have an effect on stock's price and traded volume.

The main contribution is three-fold.
First, we formalize the concept of mementum and provide a characterization based on the pairwise cointegration dynamics of (i) price and (ii) trading volumes with (iii) social media activity.
Second, based on this characterization, we use a regime-switching cointegration model to identify the meme period(s) of a stock.
The proposed method is general, that is neither stock- nor social-platform specific, and allows to infer mementum through a data-driven approach.
Finally, the methodology is applied to set of constituents of the S\&P1500, finding that the  meme stocks, so-called by newspapers, experience at least a meme period in the considered time interval.

We identify a state of the stock when the price and social volumes are cointegrated, as well as the trading volume and the social volume. Moreover, this (temporary) equilibrium characterizes a period when the coordination mechanism of investors on social media is the main driver of the dynamics of the stock's price and volumes during that window.
Conversely, several popular stocks, also characterized by intense and volatile social activity and price movements, do not exhibit meme periods.

We believe that working with pre-identified meme stocks could be misleading and limiting, as it might induce to think of ``memeness'' as an intrinsic feature of a particular stock. 
The proposed framework allows to understand whether a stock has experienced a mementum or not. Also, notice that the methodology does not require the use of a specific social platform.

\begin{figure}[t!h]
\centering
\begin{tabular}{c c}
\includegraphics[trim= 0mm 26mm 0mm 1mm,clip,height= 8cm,width= 3.3cm]{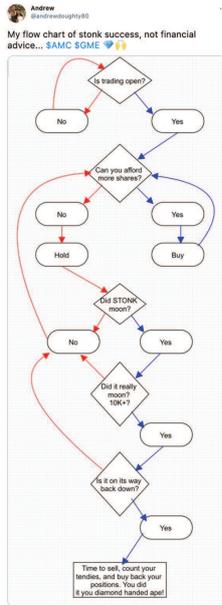} & 
\includegraphics[trim= 0mm 12mm 0mm 5mm,clip,height= 8cm]{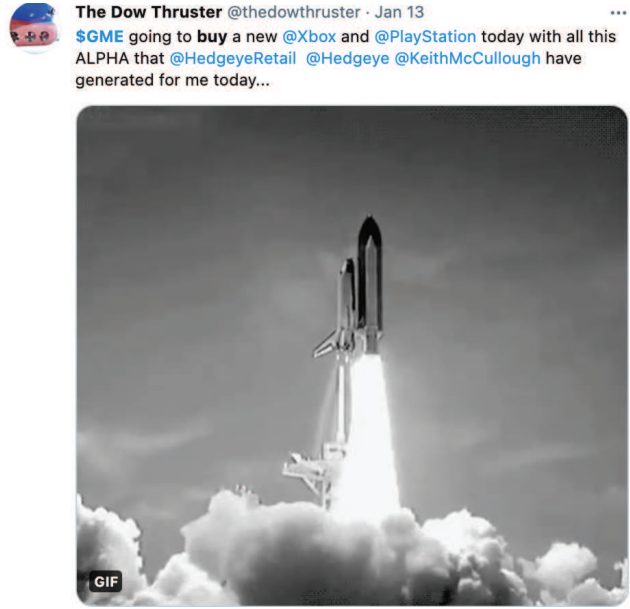} 
\end{tabular}
\caption{Examples of stock-related ``meme  tweets''.
Left: mentions GME (i.e., GameStop), AMC (i.e., AMC Entertainment), and contains an image describing the path towards ``stunk success''. Tweet link: \url{https://twitter.com/andrewdoughty80/status/1381818643789479936} 
Right: mentions GME (i.e., GameStop) and contains an image of a space rocket. Tweet link: \url{https://twitter.com/thedowthruster/status/1349398282842353666}}
\label{fig:tweet_meme}
\end{figure}

\section{On meme stocks and meme periods}

Professional and institutional investors monitor the stocks through portfolio risk managers and financial analysts' reports, conversely, many retail investors tend to ``feel the mood'' of the market by coordinating through - and participating in - online discussions on social media. Those are usually viewed as noise traders \cite{de1990noise}.
The decentralized coordination process that occurs on social media can be summarized as follows:
\begin{enumerate}
    \item a population of retail investors openly discusses on social media about popular stocks;
    \item through an assortative process, a group of retail investors that shares the belief that a given stock price can be affected by their joint purchases, coordinate on social media to synchronously buy large quantities of that stock. They do so using in their posts images and emojis (see Figure \ref{fig:tweet_meme});
    \item as the traded volumes of that stock start increasing, more investors start noticing the buying meme signals posted on social media. More investors, also driven by the fear of missing out (FOMO) start buying the stock, this creates a bandwagon effect through which stock prices and traded volumes skyrocket.
\end{enumerate}

To proxy the meme coordination activity of retail investors on social media, for each of the selected stock, we construct a daily count time-series based on posts explicitly referring to that stock. Specifically, we count the posts containing the official stock ticker symbol preceded by a tag (hash, \#, or cash, \$).\footnote{Tweets where retrieved using the FullArchive endpoint of the Twitter API V2. For each stock, we used the query: \texttt{(\$[ACRONYM] OR \#[ACRONYM]) -is:retweet has:images}. Where \texttt{[ACRONYM]} stands for the stock's official ticker symbol. \texttt{-is:retweet} is used to exclude retweets, and \texttt{has:images} is used to include only posts that contain at least one image.}
Differently from other works, which use raw daily counts of Twitter posts including both tweets and retweets \cite{umar2021tale}, we filter out retweets, that are used to share preexisting posts.

Finally, owing to the definition of \textit{meme}, that is, an image with text and/or emojis, we remove from the data all the posts that do not contain images. As a robustness check, we also build for each stock a social media time series which, in addition to the aforementioned post filtering conditions, is based only on tweets that also contain at least one emoji symbol
 related to meme stock booming.\footnote{Results obtained with these series are similar to the ones presented in this letter, and main findings remain unchanged. These results are available upon request to the authors.}
Our first choice is motivated by the fact that we want to clean from our raw series posting activity related to the imitation and diffusion of preexisting stock related contents.
As a result, we obtain a set of pruned count time-series at the daily frequency that are coherent with the concepts of ``meme stocks'' and ``mementum'', which we aim to investigate and model through this work.

The time-series plots in Figure \ref{fig:plot_time_series} show that the social media series of GameStop (GME), AMC Entertainment Holdings Inc. (AMC), and KOSS Corporation (KOSS), which have been labeled as meme stocks by multiple newspapers,\footnote{For example, see: \href{https://finance.yahoo.com/news/meme-stocks-roar-back-fueled-by-reddit-inspired-traders-191408799.html}{https://finance.yahoo.com/news/meme-stocks-roar-back-fueled-by-reddit-inspired-traders-191408799.html} and \href{https://www.reuters.com/business/retail-consumer/amc-falls-6-after-second-share-sale-this-week-2021-06-04/}{https://www.reuters.com/business/retail-consumer/amc-falls-6-after-second-share-sale-this-week-2021-06-04/}.} are characterized by abrupt spikes in mid-January 2021. Interestingly, this extreme event does not correspond to any shock to the fundamentals of a stock that may have positively perturbed its price and traded volumes. We hypothesize that this phenomenon was originated on social media, and produced a synchronized market effect both on volumes and prices. On the other side, the social media series of Pfizer (PFE), Moody's (MCO), and Disney (DIS) are more volatile and most spikes correspond to events related to the COVID-19 pandemic, which can be captured by social media activity, but clearly did not originate in social media.

\begin{figure}[t!h]
\centering
\setlength{\tabcolsep}{1.2pt}
\captionsetup{width=0.9\linewidth}
\hspace*{-10pt}
\begin{tabular}{cccc}
 & {\footnotesize Tweets} & {\footnotesize Volumes} & {\footnotesize Price} \\
\begin{rotate}{90} \hspace{25pt} {\footnotesize GME} \end{rotate} \hspace{-2pt} &
\includegraphics[trim= 15mm 0mm 15mm 0mm,clip,height= 2.6cm, width= 5.6cm]{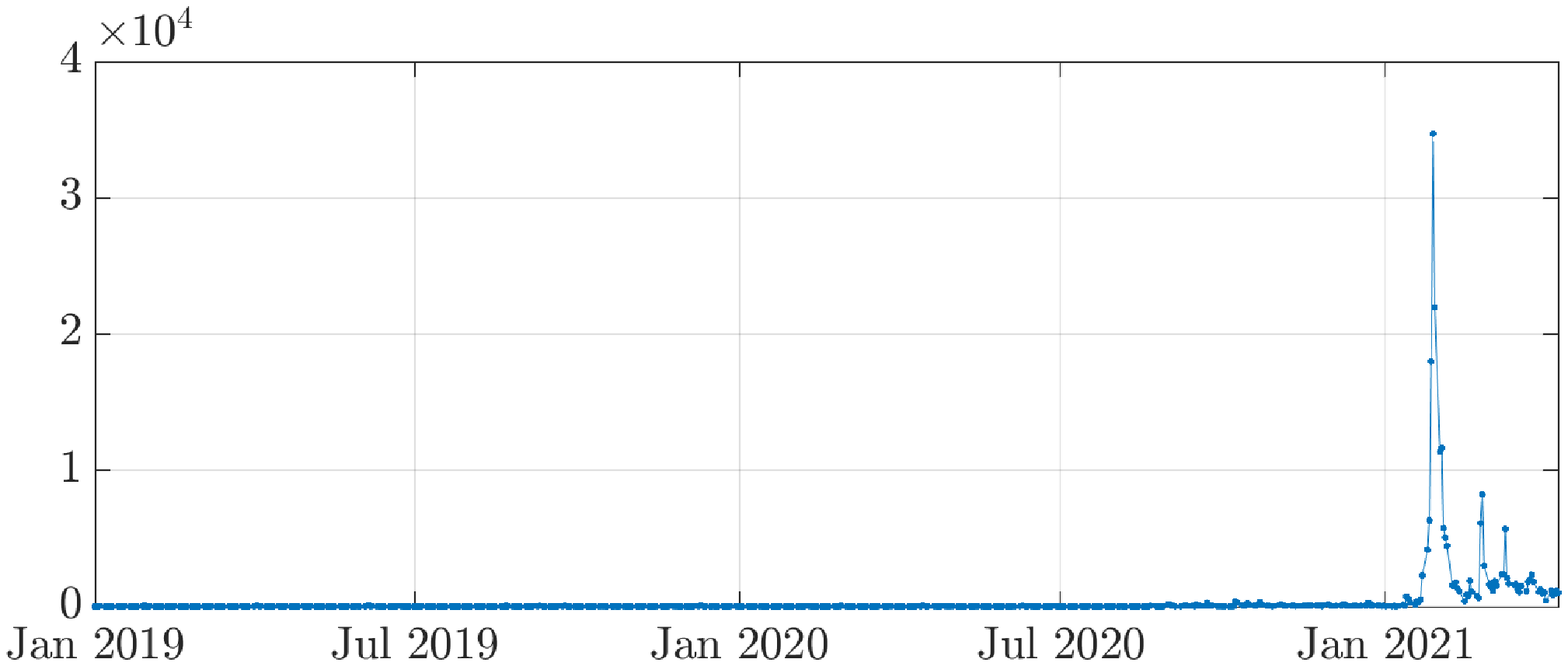} &
\includegraphics[trim= 15mm 0mm 15mm 0mm,clip,height= 2.6cm, width= 5.6cm]{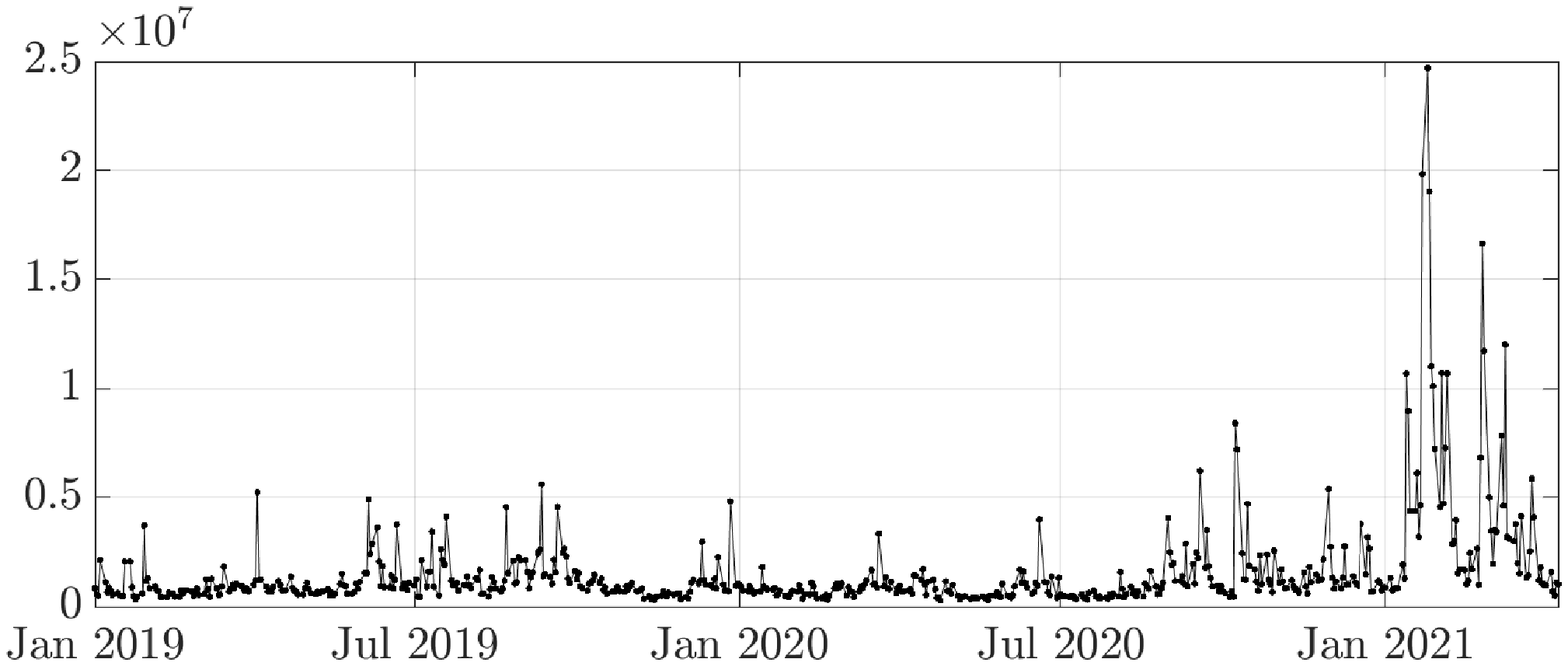} &
\includegraphics[trim= 15mm 0mm 15mm 0mm,clip,height= 2.6cm, width= 5.6cm]{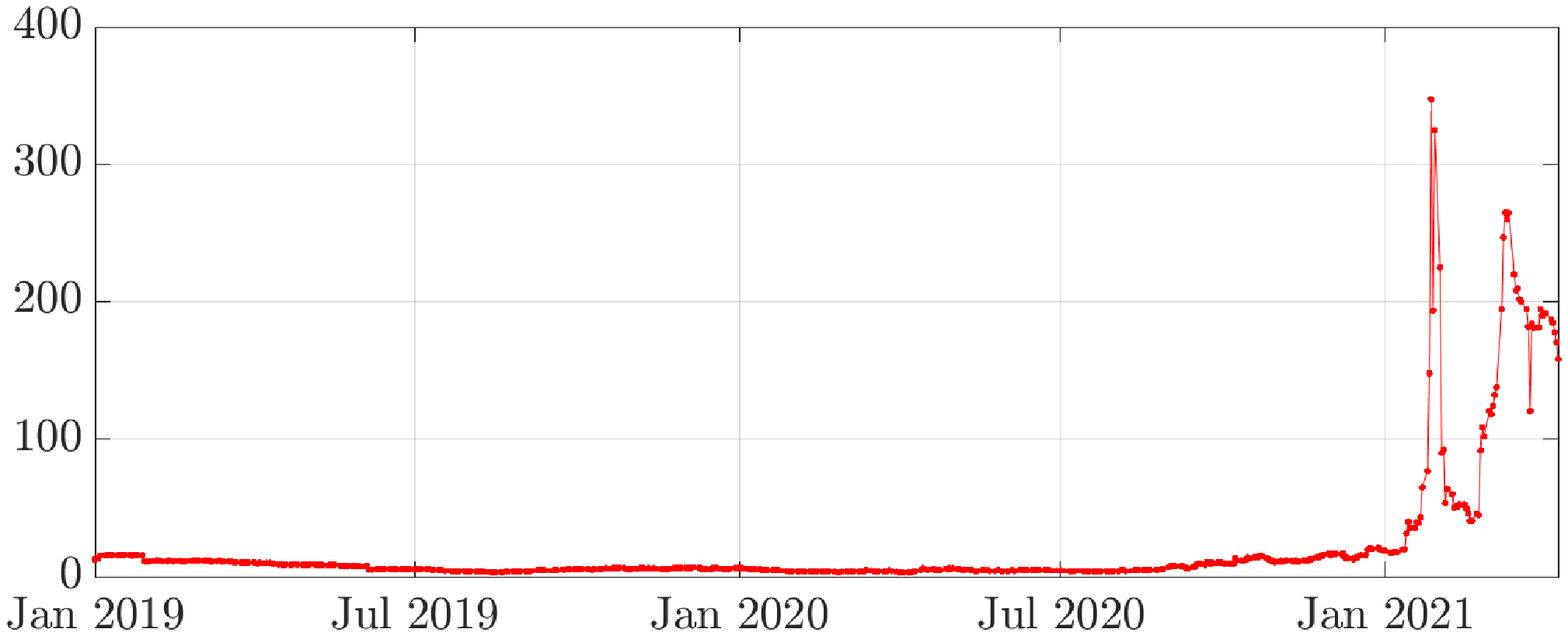} \\
\begin{rotate}{90} \hspace{25pt} {\footnotesize AMC} \end{rotate} \hspace{-2pt} &
\includegraphics[trim= 15mm 0mm 15mm 0mm,clip,height= 2.6cm, width= 5.6cm]{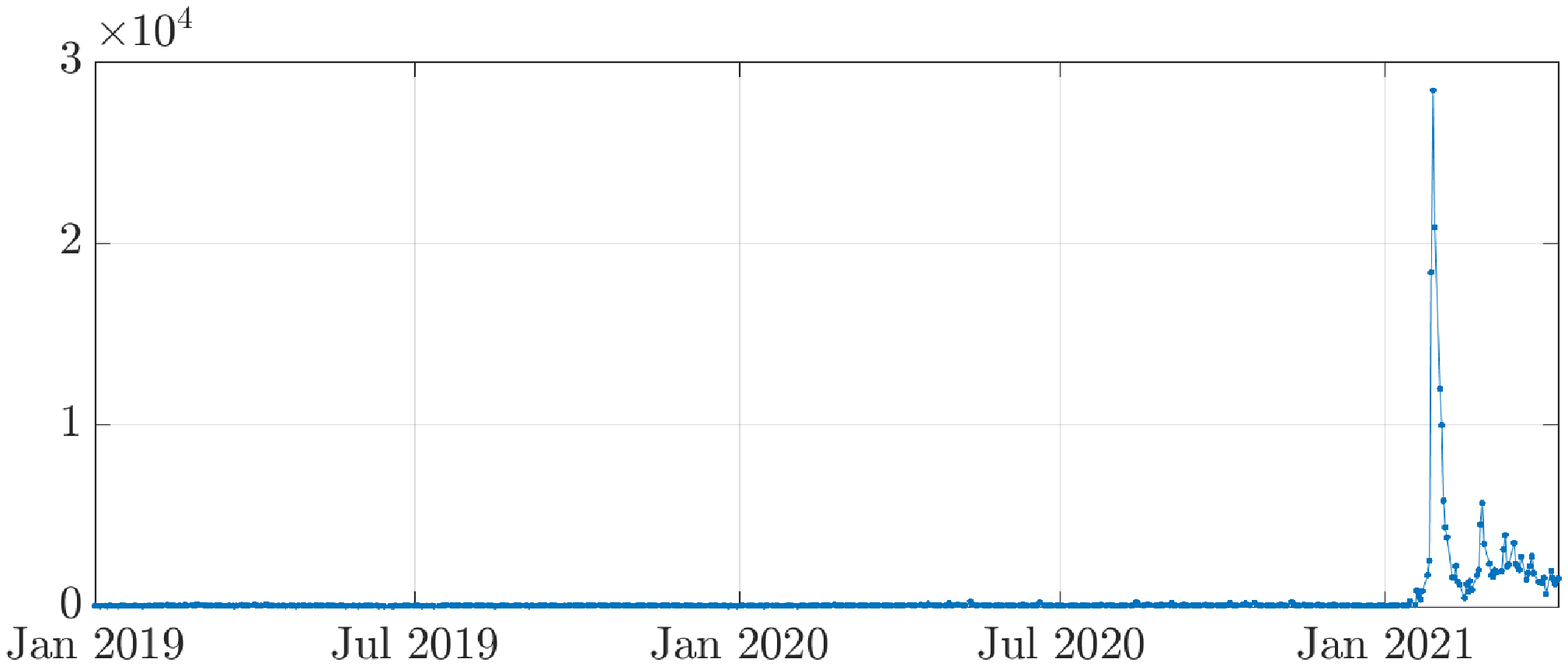} &
\includegraphics[trim= 15mm 0mm 15mm 0mm,clip,height= 2.6cm, width= 5.6cm]{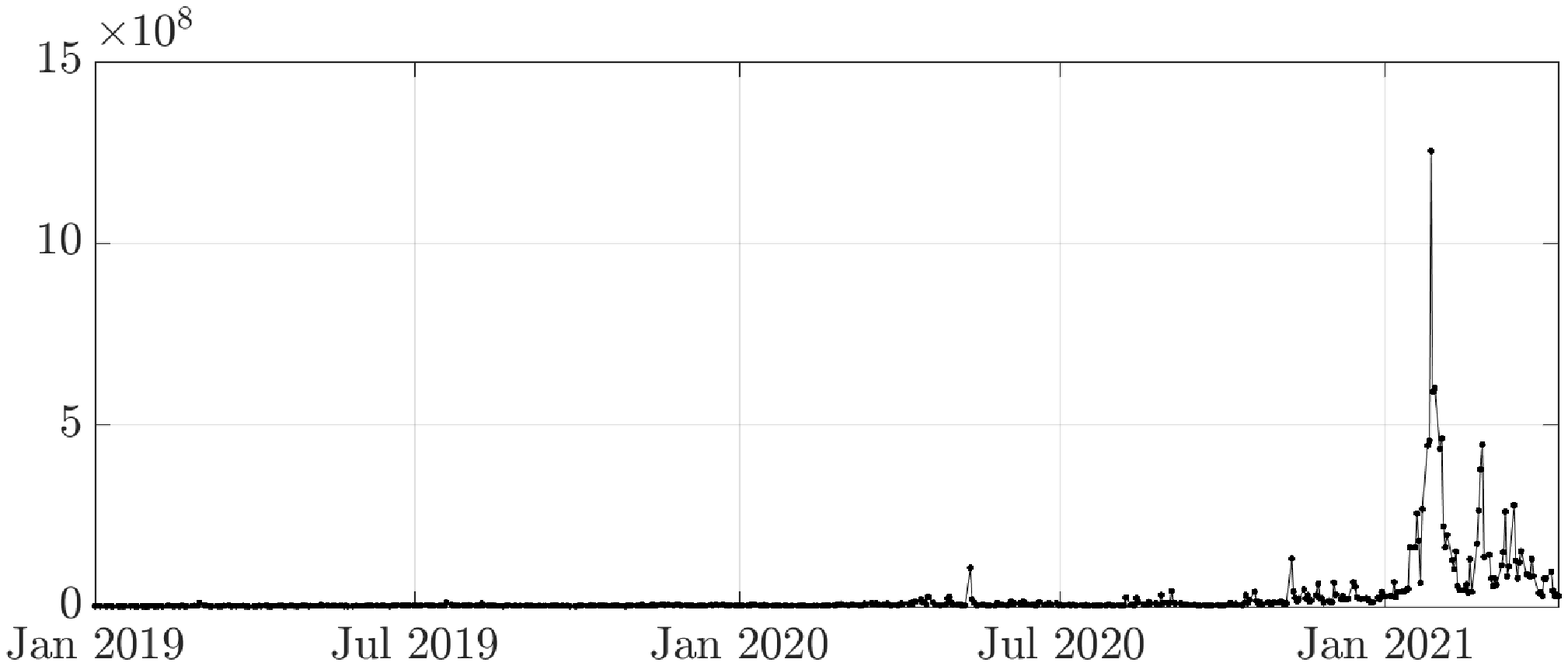} &
\includegraphics[trim= 15mm 0mm 15mm 0mm,clip,height= 2.6cm, width= 5.6cm]{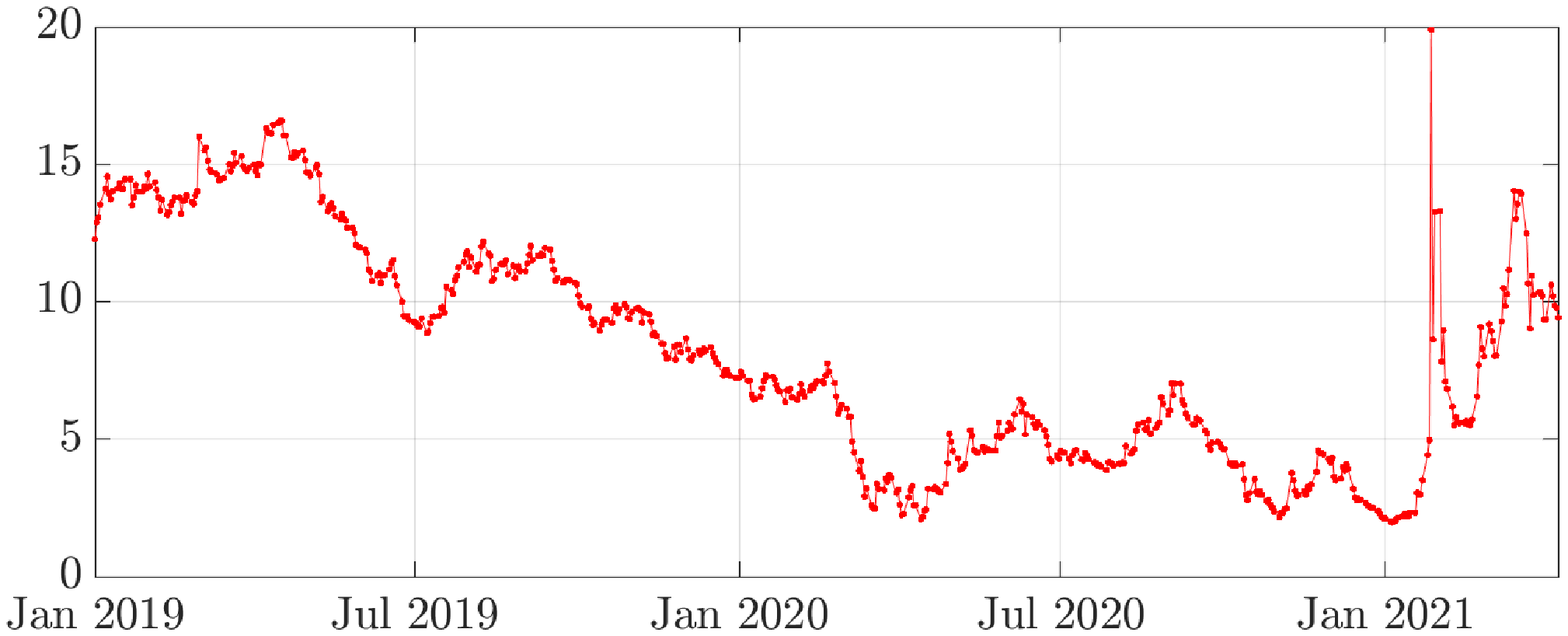} \\
\begin{rotate}{90} \hspace{25pt} {\footnotesize KOSS} \end{rotate} \hspace{-2pt} &
\includegraphics[trim= 15mm 0mm 15mm 0mm,clip,height= 2.6cm, width= 5.6cm]{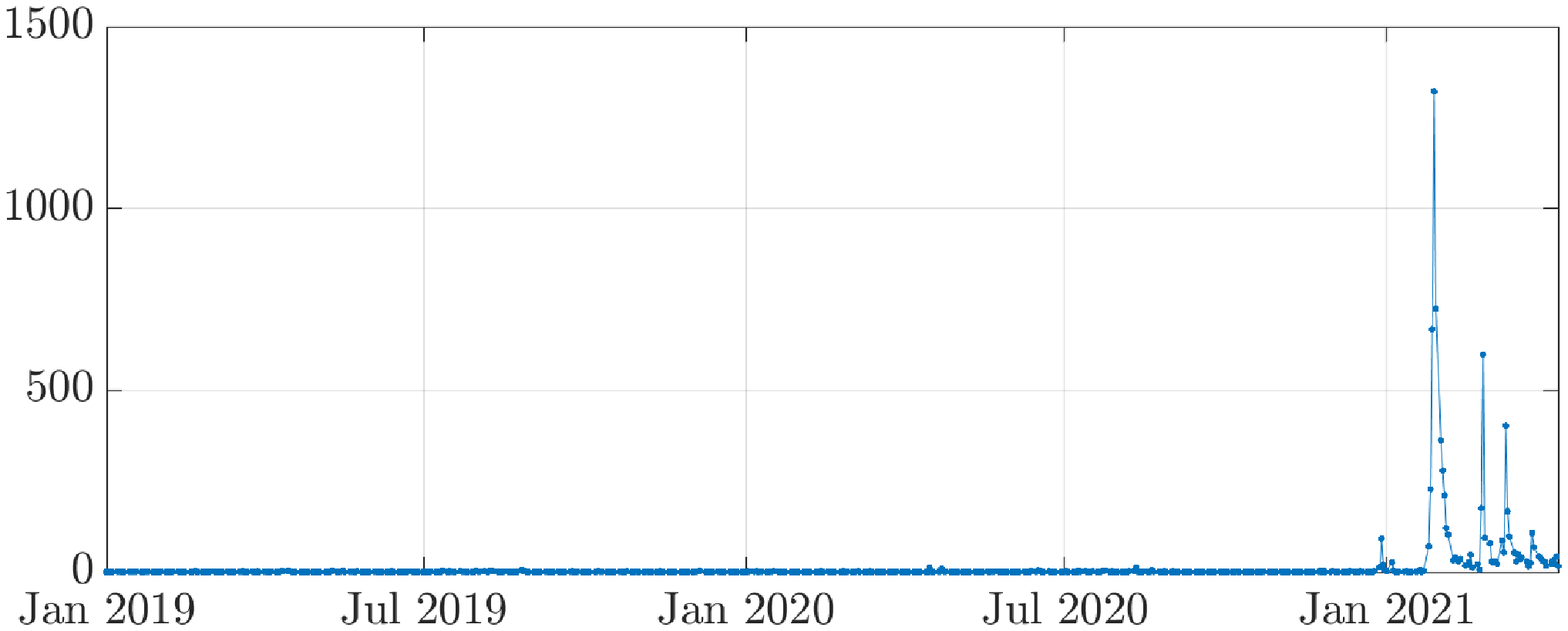} &
\includegraphics[trim= 15mm 0mm 15mm 0mm,clip,height= 2.6cm, width= 5.6cm]{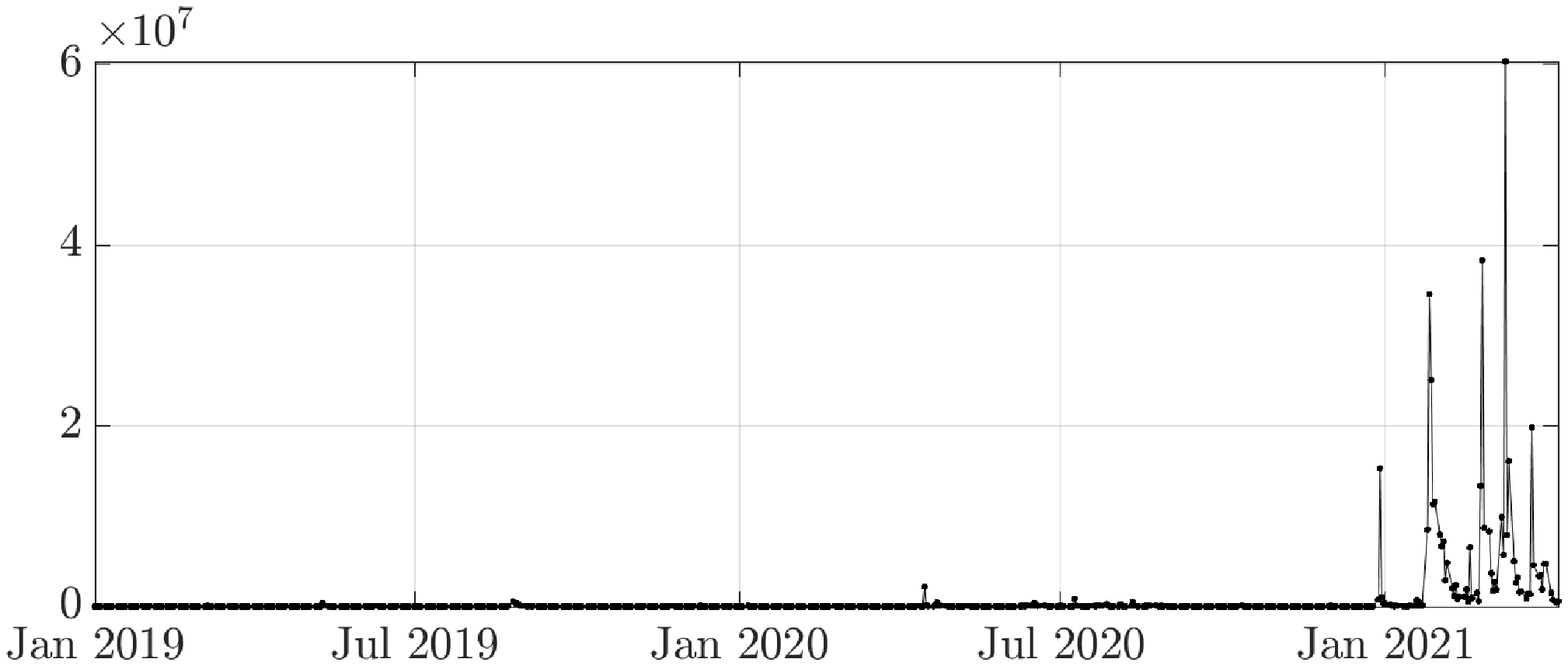} &
\includegraphics[trim= 15mm 0mm 15mm 0mm,clip,height= 2.6cm, width= 5.6cm]{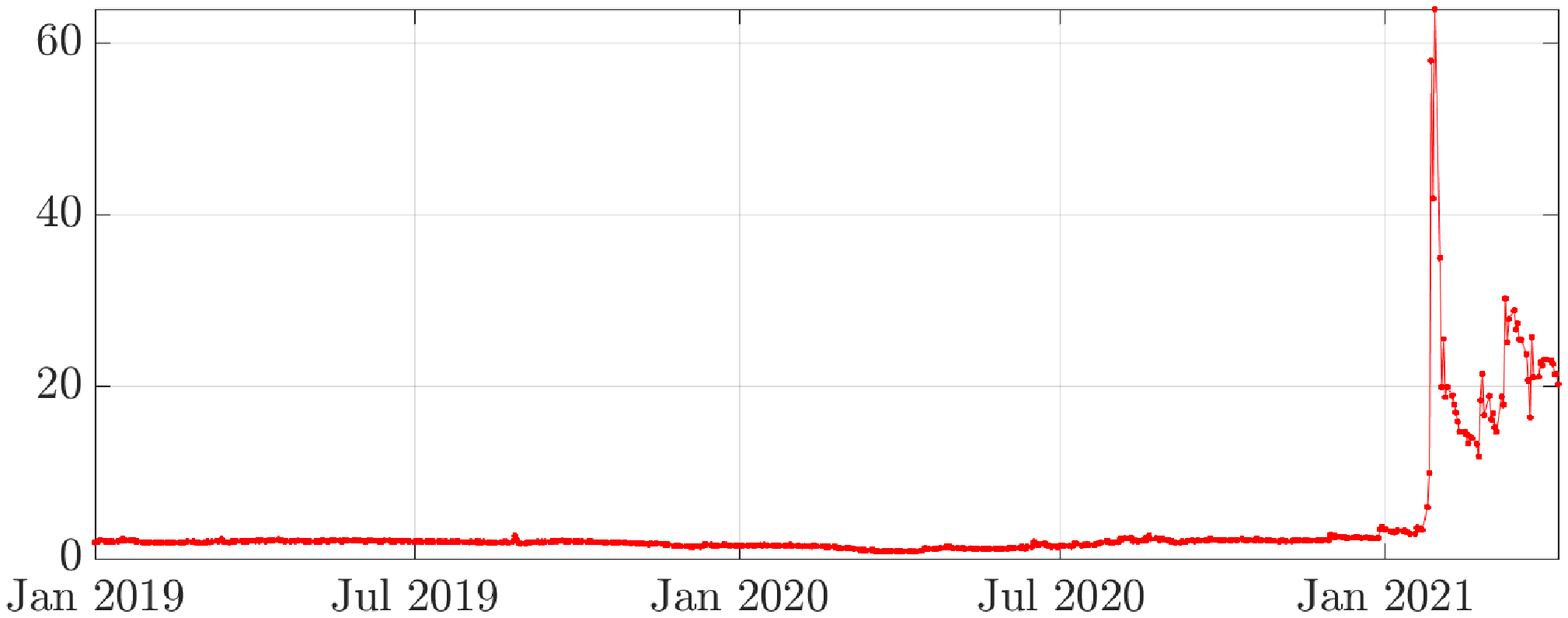} \\
\begin{rotate}{90} \hspace{25pt} {\footnotesize MCO} \end{rotate} \hspace{-2pt} &
\includegraphics[trim= 15mm 0mm 15mm 0mm,clip,height= 2.6cm, width= 5.6cm]{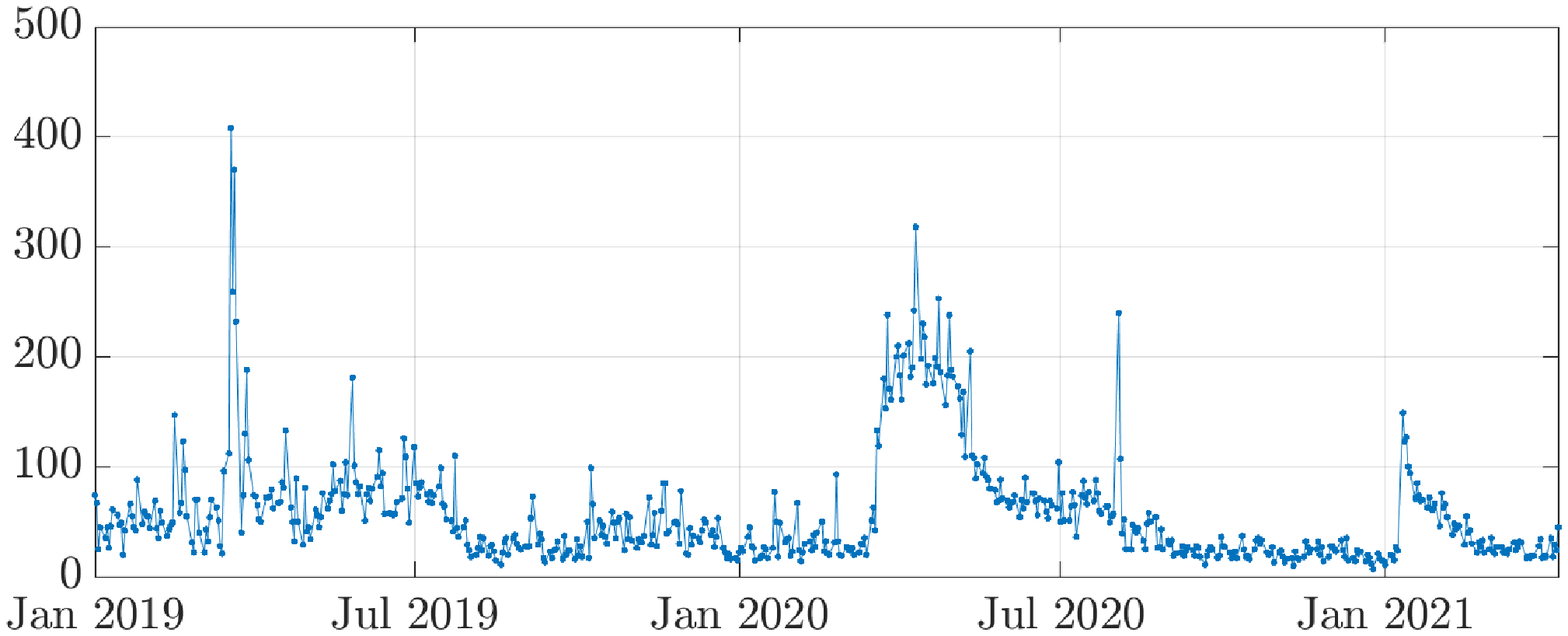} &
\includegraphics[trim= 15mm 0mm 15mm 0mm,clip,height= 2.6cm, width= 5.6cm]{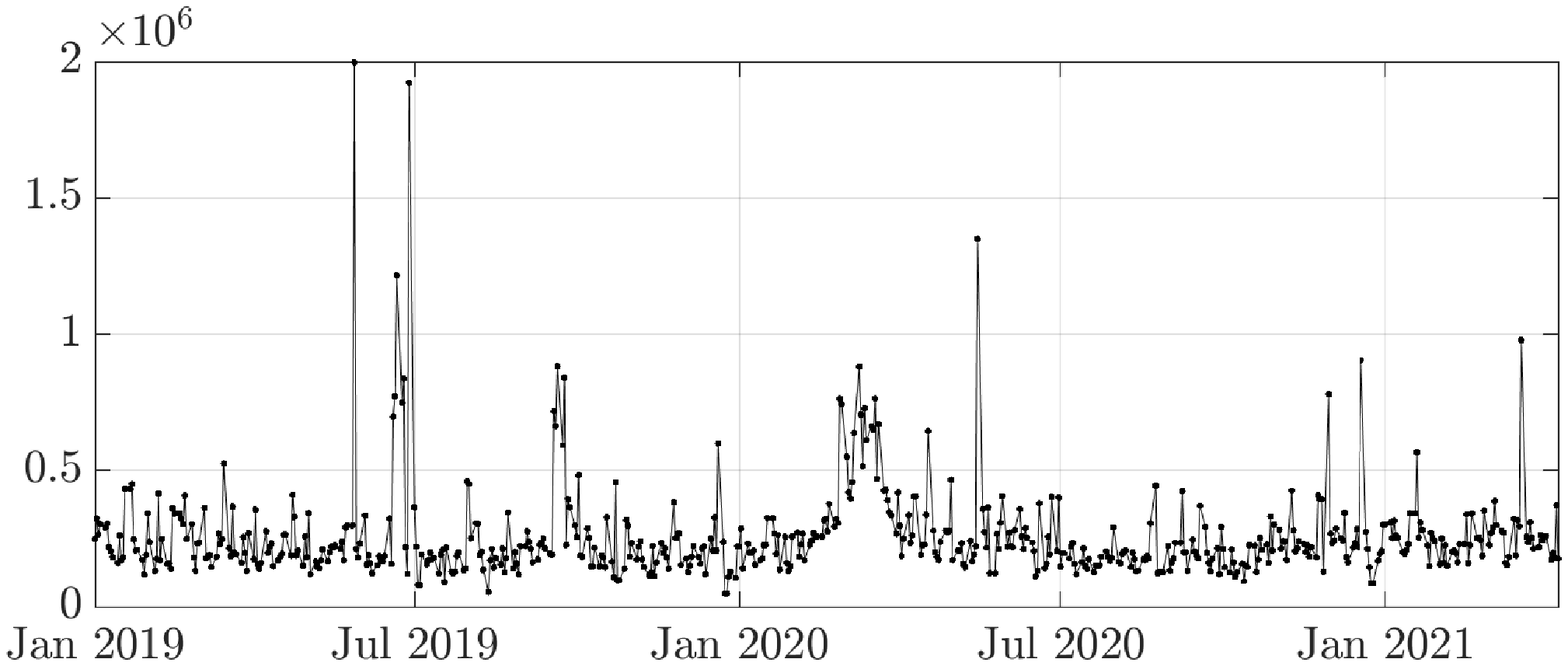} &
\includegraphics[trim= 15mm 0mm 15mm 0mm,clip,height= 2.6cm, width= 5.6cm]{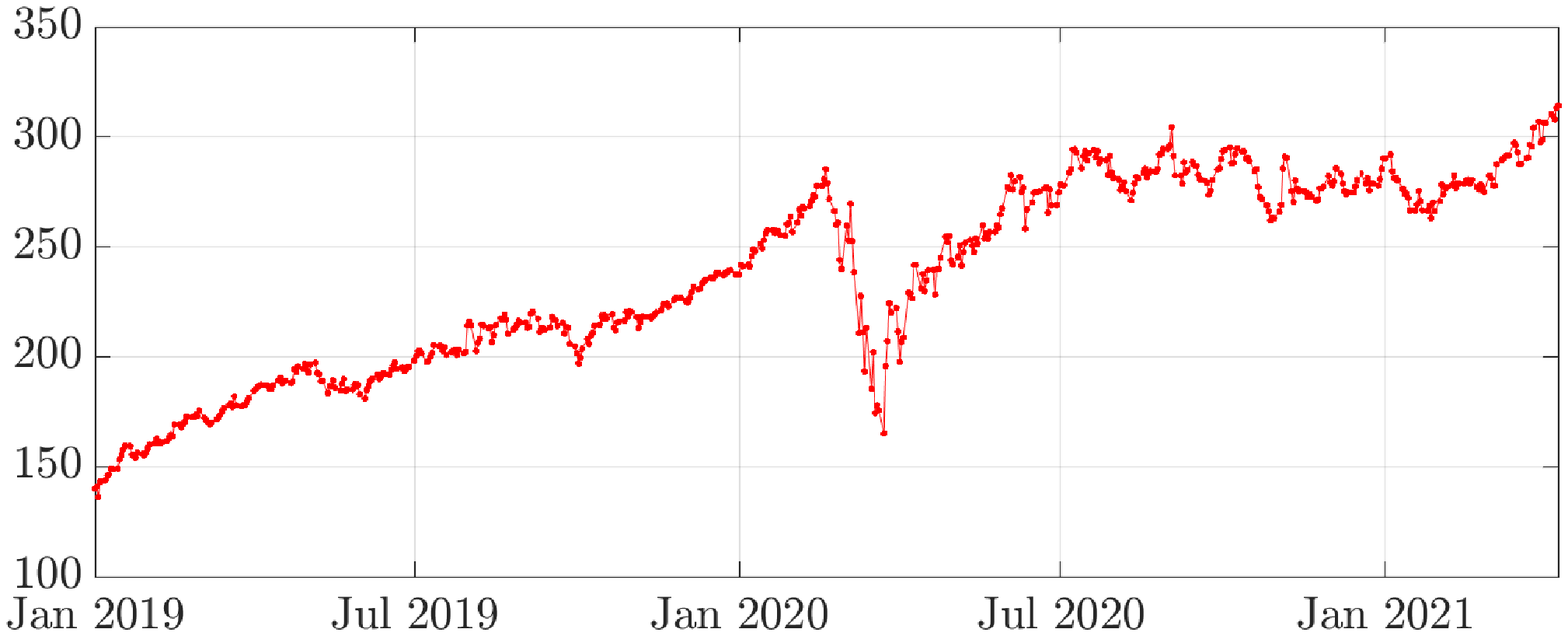} \\
\begin{rotate}{90} \hspace{25pt} {\footnotesize PFE} \end{rotate} \hspace{-2pt} &
\includegraphics[trim= 15mm 0mm 15mm 0mm,clip,height= 2.6cm, width= 5.6cm]{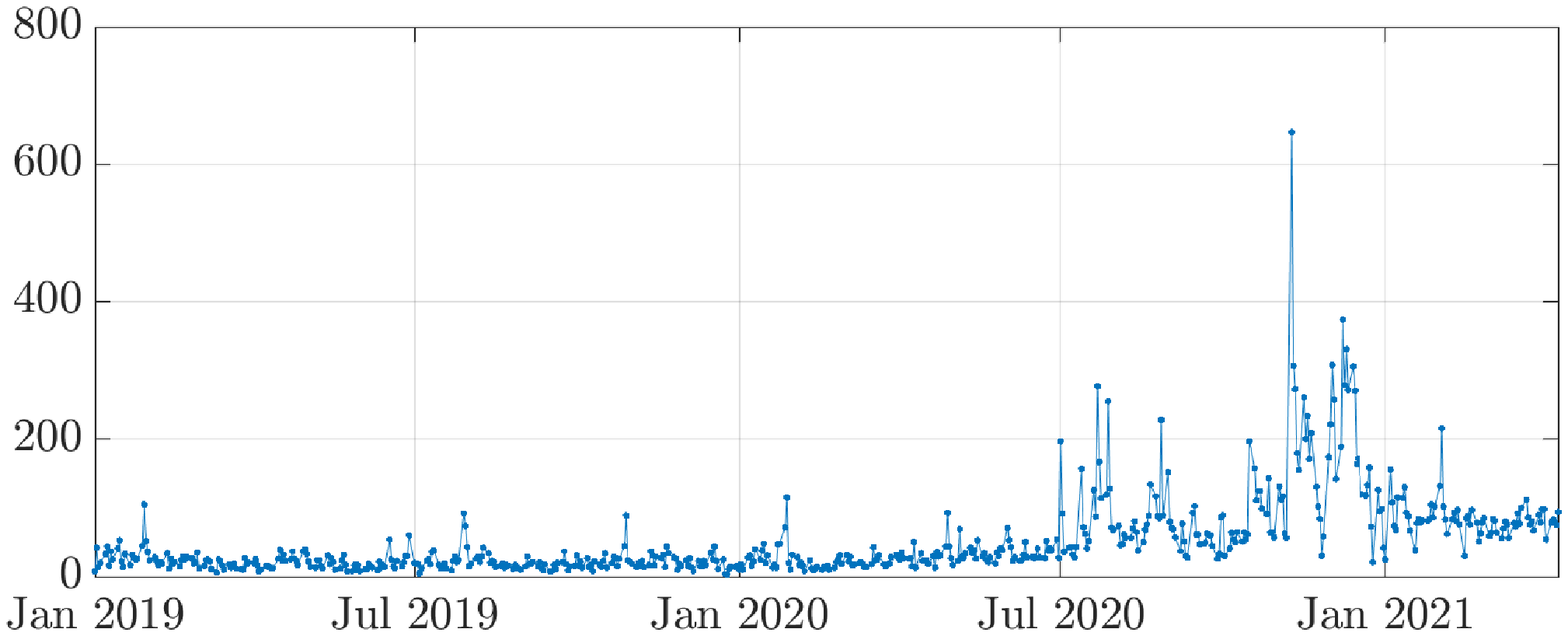} &
\includegraphics[trim= 15mm 0mm 15mm 0mm,clip,height= 2.6cm, width= 5.6cm]{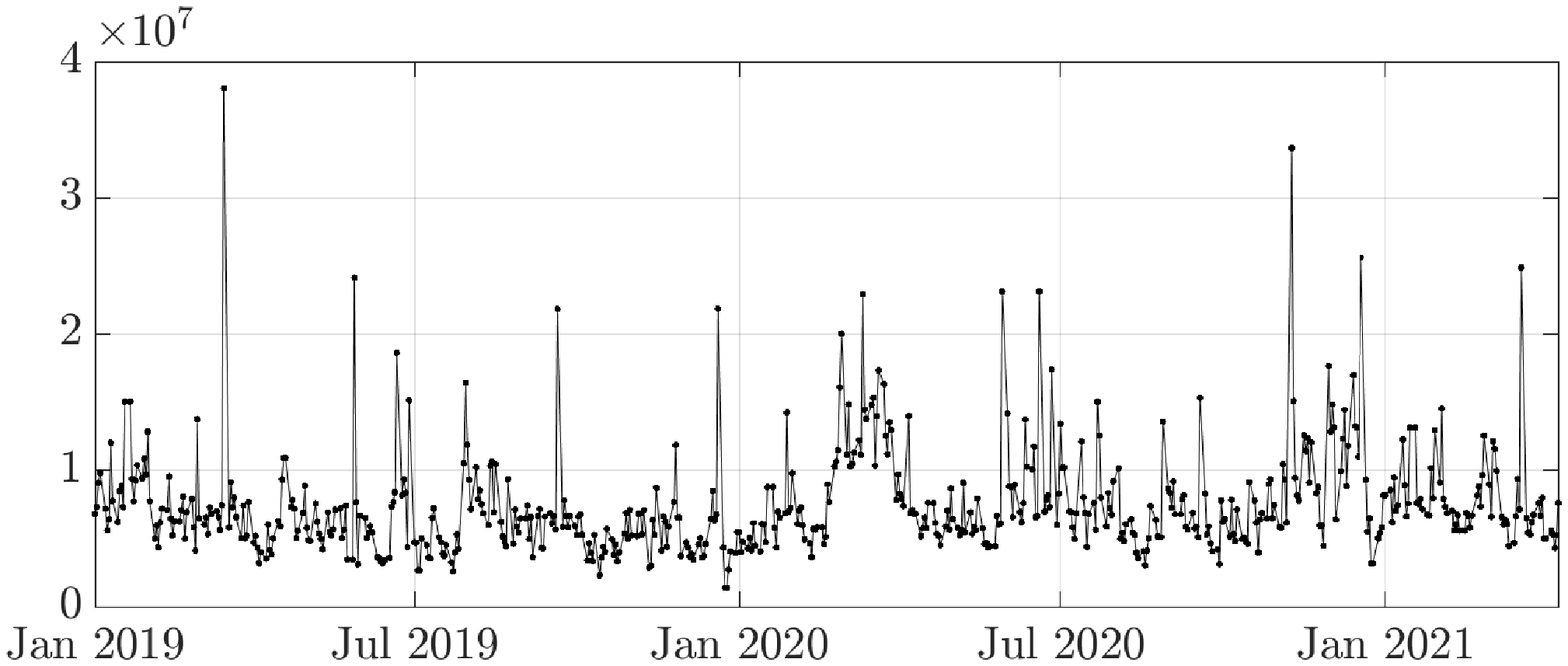} &
\includegraphics[trim= 15mm 0mm 15mm 0mm,clip,height= 2.6cm, width= 5.6cm]{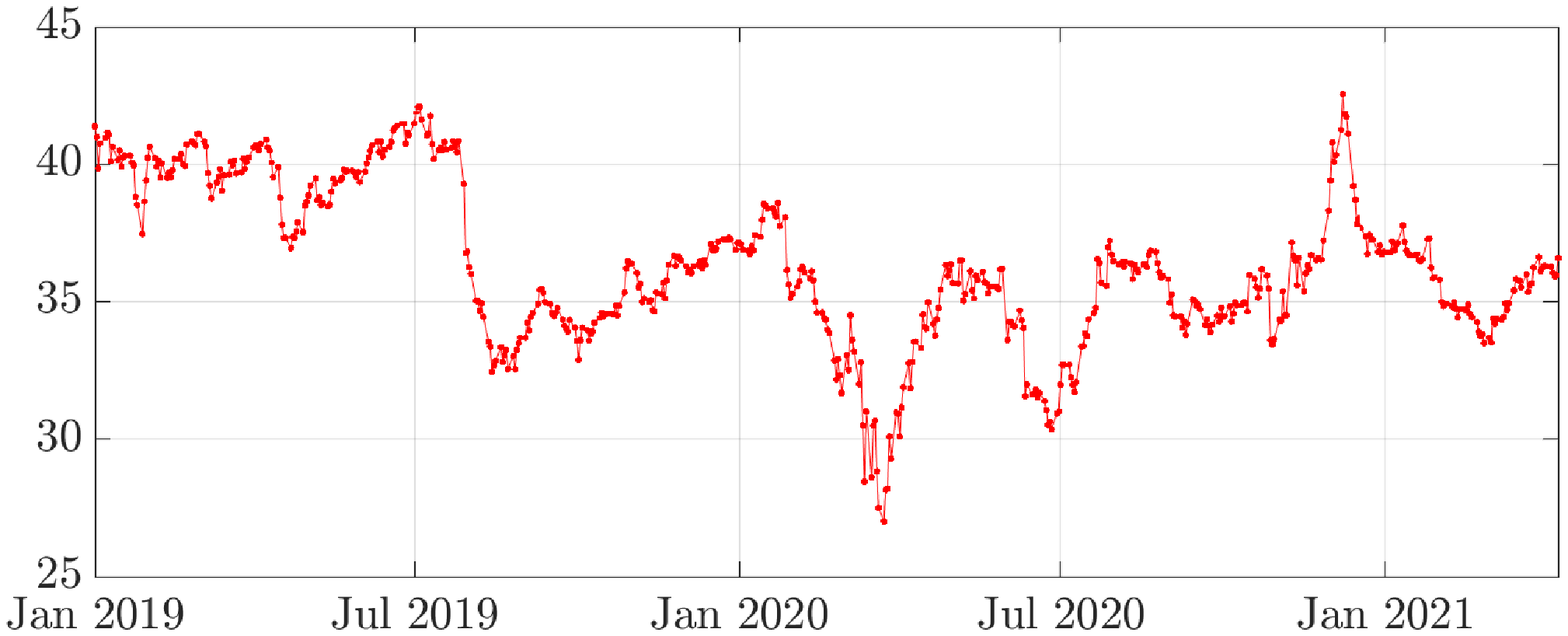} \\
\begin{rotate}{90} \hspace{25pt} {\footnotesize DIS} \end{rotate} \hspace{-2pt} &
\includegraphics[trim= 15mm 0mm 15mm 0mm,clip,height= 2.6cm, width= 5.6cm]{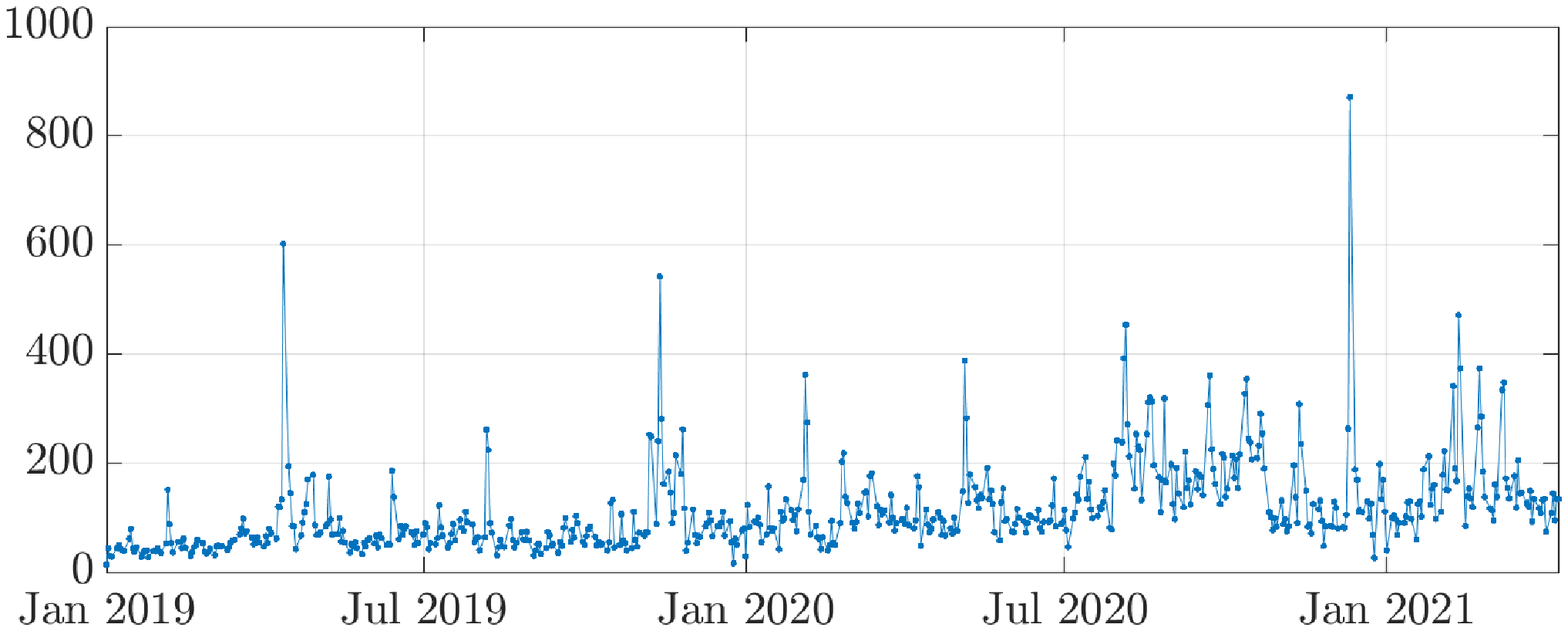} &
\includegraphics[trim= 15mm 0mm 15mm 0mm,clip,height= 2.6cm, width= 5.6cm]{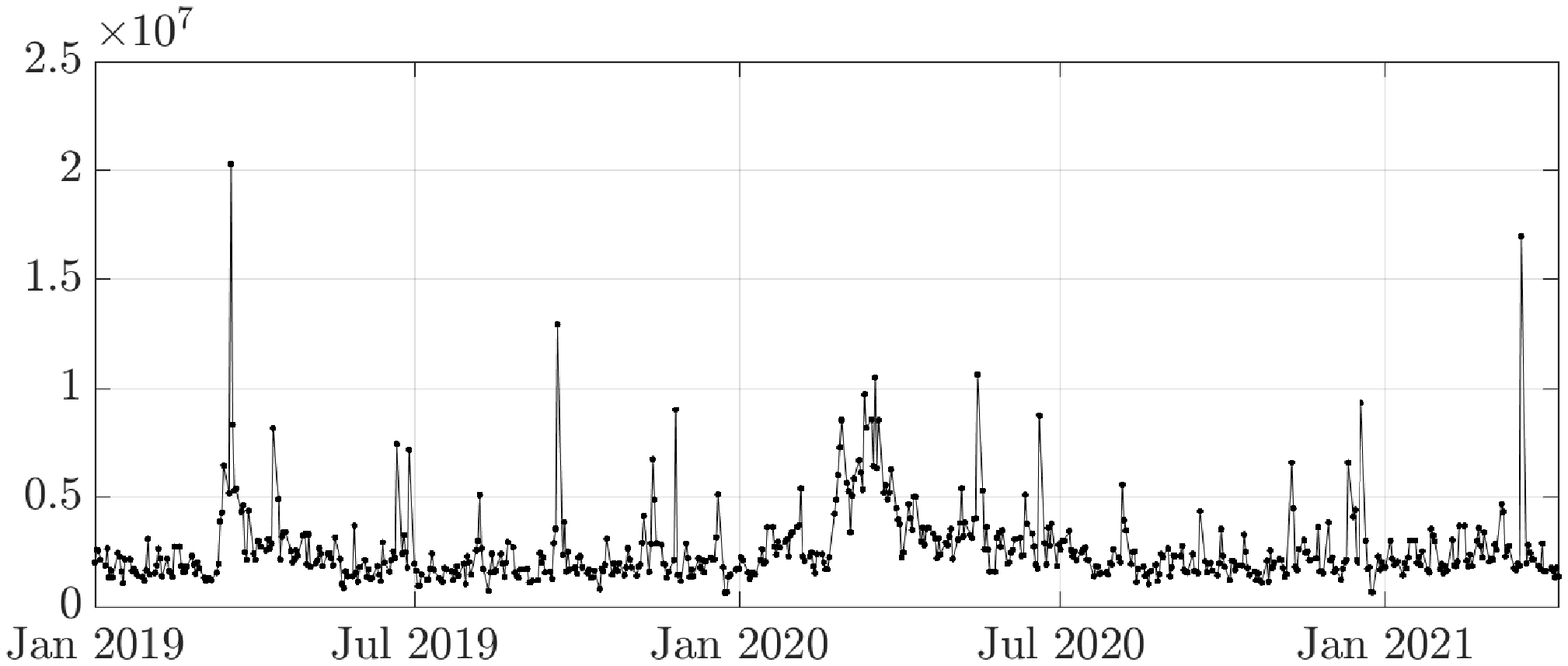} &
\includegraphics[trim= 15mm 0mm 15mm 0mm,clip,height= 2.6cm, width= 5.6cm]{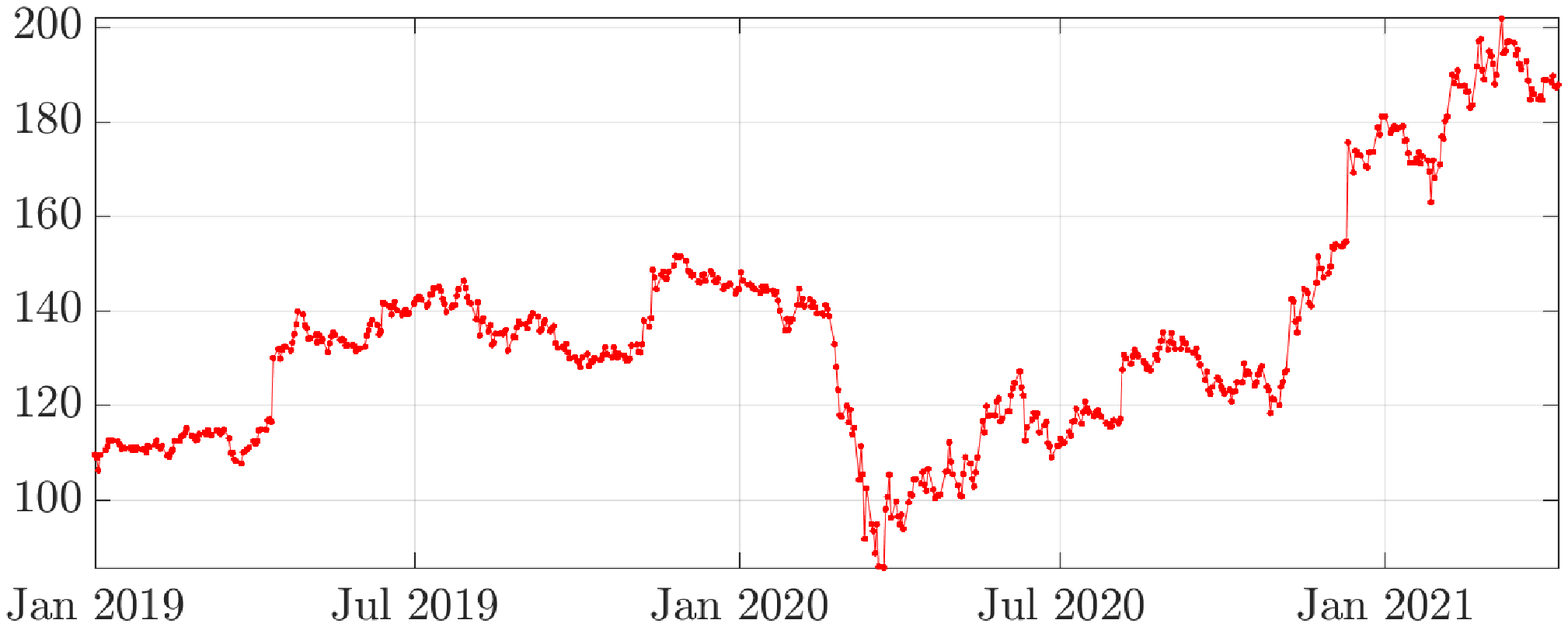} 
\end{tabular}
\caption{Time series of tweet counts (first column), traded volume (second column), and price (third column) for GameStop (GME), AMC Entertainment (AMC), KOSS Corporation (KOSS), Moody's (MCO), Pfizer (PFE), and Disney (DIS).}
\label{fig:plot_time_series}
\end{figure}

We now provide a characterization of the concept of ``\textit{mementum}'' of a stock that relies on the properties of its social and financial time series. We focus on Twitter, because it is a popular and generalist social platform, which provides an API for the free download of data. However, the following definition and procedure are still valid if other social media are considered.

First, we postulate the contemporaneous existence of a relationship between the series of prices and Twitter posts with images, as well as between the latter and trading volumes.
This reflects the effects of buying coordination through social platforms, which, increasing the demand for a stock, is expected to positively affect both prices and traded volumes of the concerned stocks.
Second, since these two relationships co-exist during a meme phenomenon, we require them to be synchronized in time. Moreover, as these coordination mechanisms cause structural changes of the social and financial series, we require the above-mentioned relationships to hold after a regime-switch.
Finally, to account for the persistence of the meme phenomenon, we impose a  minimum duration condition that removes isolated and short-lived events.
The following definition summarizes these points.

\begin{definition}[Mementum] \label{def:meme_period}
We say that a given stock experiences a \textit{mementum} (i.e., a meme period) if there exists a non-empty temporal window such that the following conditions jointly hold:
\begin{itemize}
\item \textit{Condition 1:} there is syncronicity in cointegration between (i) price and tweet series; and (ii) volumes and tweet series;
\item  \textit{Condition 2:} there is syncronicity in the timing when the regime switches to cointegration;
\item  \textit{Condition 3:}  for both series, the cointegration regime is persistent, as well as the regime prior to cointegration.
\end{itemize}
\end{definition}

\section{Procedure for identifying the ``mementum''}

Cointegration among $n$ time series (either I(0) or I(1)) is defined as the presence of $r < n$ stationary linear combinations of them \cite{engle1987co,johansen1991estimation}. In the standard static setting, each relationship can be interpreted as a long-run relationship and $r$, the number of such relationships, is called the cointegration rank. The static cointegration framework allows to detect log-run relationships among variables, but fails to account for structural changes and time variation of them, which is indeed of paramount importance in characterizing meme periods (see Definition~\ref{def:meme_period}).
To deal with this issue, we need a time-varying cointegrating framework, which enables the identification of dynamic price-tweet and volume-tweet pairwise cointegrating relationships.

Recently, \cite{chua2018bayesian} proposed an econometric framework that allows for both a time-varying cointegrating matrix and a time-varying cointegrating rank. In this setting, a cointegration relationship represents a smoothly changing equilibrium towards which the variables are attracted at a specific point in time, but not necessarily at each point.
The time-varying rank and parameters vector error correction model (TVR-TVP-VECM) to study cointegration is given by
\begin{equation}
\Delta y_t = c + y_{t-1} \Pi_t + \Delta y_{t-1}B + \epsilon_t,  \qquad  \epsilon_t \distas{iid} \mathcal{N}(0,\Sigma),
\label{eq:model}
\end{equation}
where $y_t = (y_{1,t},\ldots,y_{n,t})$, $\Delta$ is the first difference operator, and $\Pi_t$ is the $(n \times n)$ time-varying cointegrating matrix. If at time $t$ there exists a non-empty set of cointegrating relationships, then $\Pi_t$ has rank $r_t < n$ and admits a low rank decomposition as $\Pi_t = \boldsymbol{\beta}_t \boldsymbol{\alpha}_t$, where $\boldsymbol{\beta}_t$ and $\boldsymbol{\alpha}_t$ are $(n \times r_t)$ and $(r_t \times n)$ matrices of cointegrating relationships and loadings, respectively.
Specifically, they allow the rows of the loading matrix, $\boldsymbol{\alpha}$, and the columns of the cointegrating relationships matrix, $\boldsymbol{\beta}$, to vary smoothly over time (similar to the TVP-VECM setting of \cite{koop2011bayesian}), independently of each other.
Finally, the dynamics of the rank is driven by a homogeneous $N$-state hidden Markov chain $S_t$ with transition matrix $P$, with entries $p_{ij} = \mathbb{P}(S_t = j | S_{t-1} = i)$, $i,j=1,\ldots,n+1$. Defining the indicator $s_{jt} = 1$ if $S_t = j$, the dynamics of $\Pi_t$ is given by
\begin{equation}
\Pi_t = U_t I(S_t) I(S_t) \Lambda_t V_t' = U_t \kappa_t I(S_t) I(S_t) \kappa_t^{-1} \Lambda_t V_t' 
\label{eq:model_Pi}
\end{equation}
where $U_t,V_t,\Lambda_t$ stem from the SVD decomposition of $\Pi_t$, with $U_t,V_t$ are orthogonal matrices and $\Lambda_t$ is diagonal, $\kappa_t$ is an auxiliary diagonal matrix, and $I(S_t)$ is a regime-dependent diagonal matrix with $i$-th diagonal element $I(S_t)_{ii} = (1-s_{1t})\sum_{j=i+1}^{n+1} s_{jt}$.
The assumption of a Markovian process for $r_t$ allows for persistence of the time-varying rank, which is a desirable property in our framework. Notice that, at each point in time the rank of $\Pi_t$ is fully determined by the chain $S_t$.
%

Summarizing, this model allows for alternating cointegrated and noncointegrated states, as well as time-variation in the cointegrating parameters.
This enables the variables to exhibit a common stochastic trend only for specific periods, which is in line with some of the features of mementum, as in Definition~\ref{def:meme_period}.
We refer to \cite{chua2018bayesian} for further details about the properties of the model and the inference procedure.
The estimation is performed according to a Bayesian approach relying on a Markov chain Monte Carlo algorithm to approximate the joint posterior distribution.\footnote{The estimation took about 10 minutes on a 3.1GHz laptop to obtain 5,000 draws.}

Since we are interested in pairwise cointegrating relationships, for each stock, we estimate model \eqref{eq:model} for the two pairs pr/tw (price and tweets) and vol/tw (volumes and tweets).
The estimated regimes allow us to identify the periods where both pairs of variables are cointegrated. This is used as input to the following procedure for identifying the meme periods.

For each stock, we take all starting dates of price-tweet and volume-tweet cointegration regimes, retaining only those that last at least $d_c$ days (\textit{Condition 3}).
To reduce the influence of noisy estimates, we impose a further persistence condition. Specifically, we filter out all dates that were not preceded by at least $d_p$ days of a non-cointegrated regime.
For the same reason, we ignore extremely transitory falls (of duration $\leq d_f$) from the cointegration regime, by merging  cointegrated intervals that closely follow each other. 
Hence, we match the starting dates of price-tweet and volume-tweet cointegrated regimes (\textit{Condition 2}). We allow for $d_w$ days of delay. Once matched we keep only the intersection of dates between the two sets of time-intervals (\textit{Condition 1}), keeping, as before, only jointly cointegrated periods that last at least $d_c$ days (\textit{Condition 3}).

\begin{example}
\begin{figure}[t!h]
\centering
\includegraphics[trim= 0mm 0mm 0mm 0mm,clip, width= \textwidth]{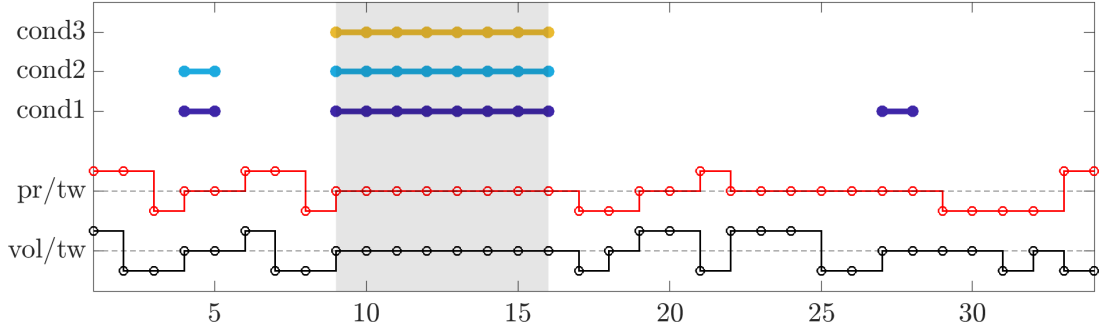}
\caption{Example of the method on simulated data. Estimated cointegration regimes for the pairs price/tweets (red line) and volumes/tweets (black line). The dashed lines represent the regime with 1 cointegration relationship. The colored bars in the top three lines identify the periods when the three conditions in Definition~\ref{def:meme_period} are satisfied (condition 1 in dark blue, 2 in light blue, 3 in dark yellow).}
\label{fig:example}
\end{figure}
\FloatBarrier
Figure~\ref{fig:example} provides a graphical representation of the procedure for a simulated example.
The solid lines on the bottom report the estimated regimes for the two pairs of series (pr/tw and vol/tw), with a dashed line indicating the regime corresponding to a cointegration relationship (rank 1). The top three row use color shadings to denote the days when the corresponding condition in Definition~\ref{def:meme_period} is satisfied.
A meme period starts when all the conditions are satisfied, at $t=9$, and lasts until $t=16$, when at least one condition is violated. Notice that at $t=4$ conditions 3 is violated (cointegrating relationships are not sufficiently persistent), whereas at $t=27$ only condition 1 holds.
\end{example}

\begin{figure}[th]
\centering
\setlength{\abovecaptionskip}{-3pt}
\setlength{\tabcolsep}{-10pt}
\footnotesize
\captionsetup{width=0.9\linewidth}
\begin{tabular}{c C{0.72\textwidth} >{\centering\arraybackslash}m{0.2\textwidth}}
 & {\footnotesize Cointegrated regimes} & {\footnotesize Mementum} \\
\begin{rotate}{90} \hspace{36pt} {\footnotesize GME} \end{rotate} \hspace{-7pt} &
\includegraphics[trim= 5mm 0mm 1mm 0mm,clip,height= 3.1cm, width= 10.2cm]{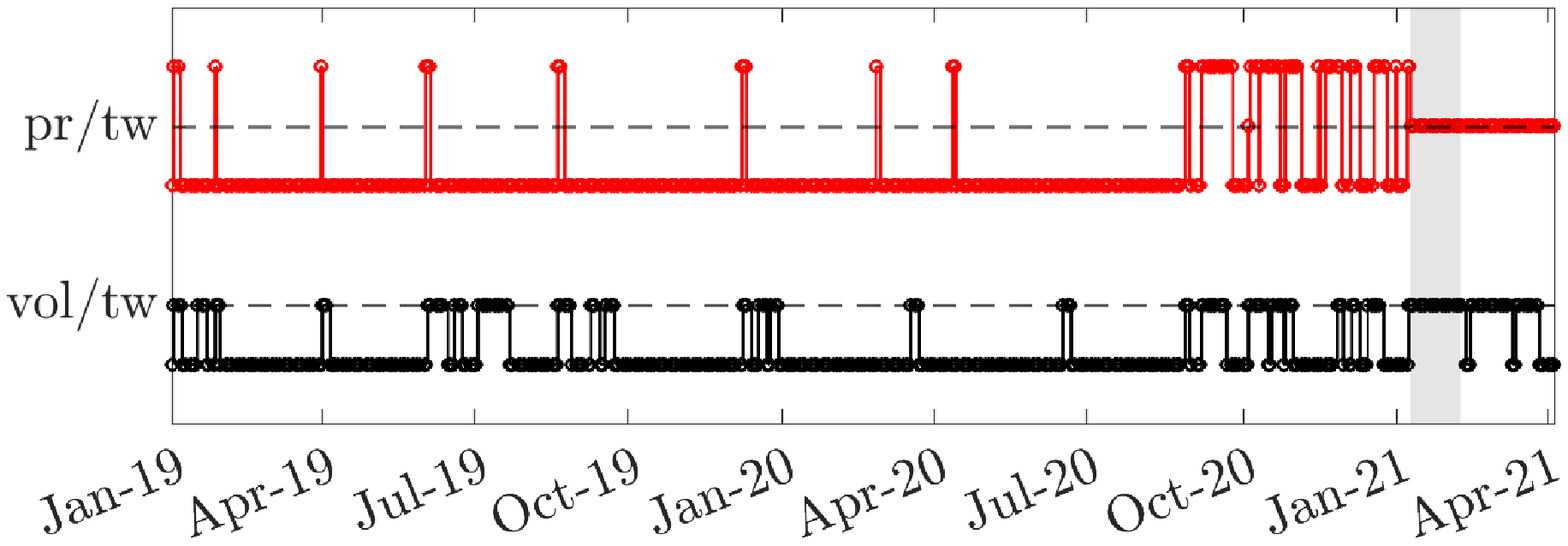} &
\vspace{-10ex}
\begin{minipage}[c]{0.25\textwidth}
\footnotesize
13 Jan 2020 -- 12 Feb 2021
\end{minipage} \vspace{10ex} \\[-5pt]
\begin{rotate}{90} \hspace{36pt} {\footnotesize AMC} \end{rotate} \hspace{-7pt} &
\includegraphics[trim= 5mm 0mm 1mm 0mm,clip,height= 3.1cm, width= 10.2cm]{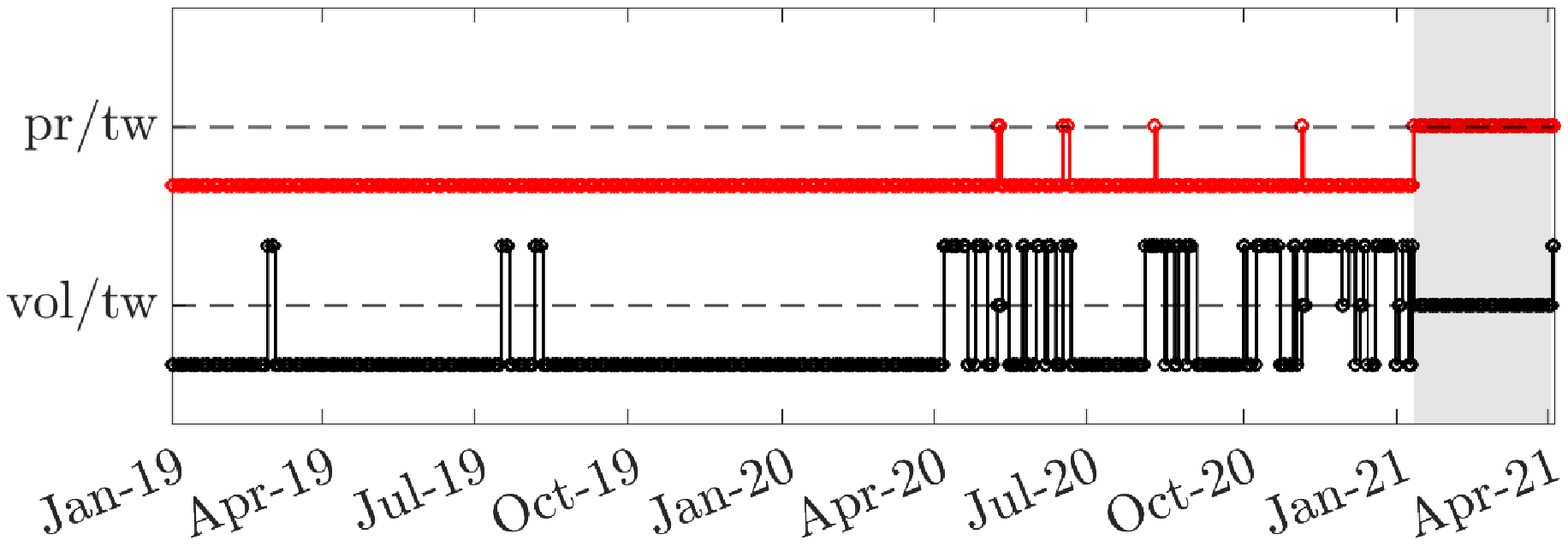} &
\vspace{-10ex}
\begin{minipage}[c]{0.3\textwidth}
\footnotesize
15 Jan 2020 -- 7 Apr 2021
\end{minipage} \vspace{10ex} \\[-5pt]
\begin{rotate}{90} \hspace{36pt} {\footnotesize KOSS} \end{rotate} \hspace{-7pt} &
\includegraphics[trim= 5mm 0mm 1mm 0mm,clip,height= 3.1cm, width= 10.2cm]{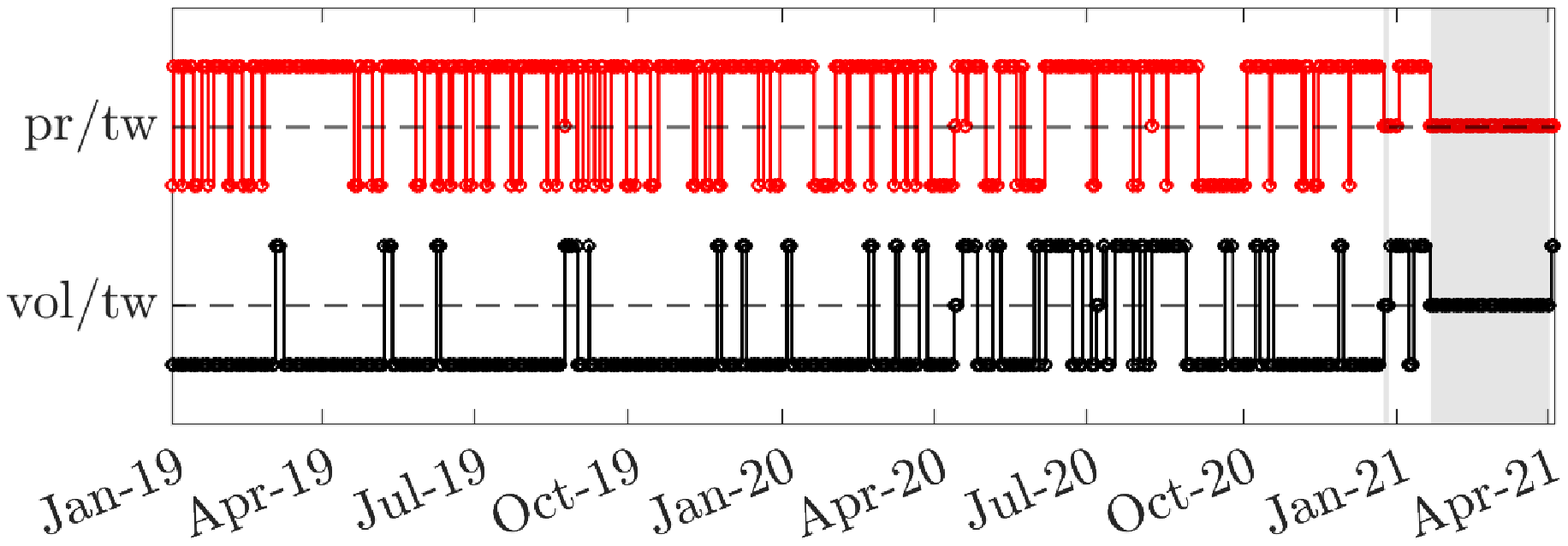} &
\vspace{-10ex}
\begin{minipage}[c]{0.3\textwidth}
\footnotesize
28 Dec 2020 -- 31 Dec 2021,\\ 25 Jan 2020 -- 6 Apr 2021
\end{minipage} \vspace{10ex} \\[-5pt]
\begin{rotate}{90} \hspace{36pt} {\footnotesize MCO} \end{rotate} \hspace{-7pt} &
\includegraphics[trim= 5mm 0mm 1mm 0mm,clip,height= 3.1cm, width= 10.2cm]{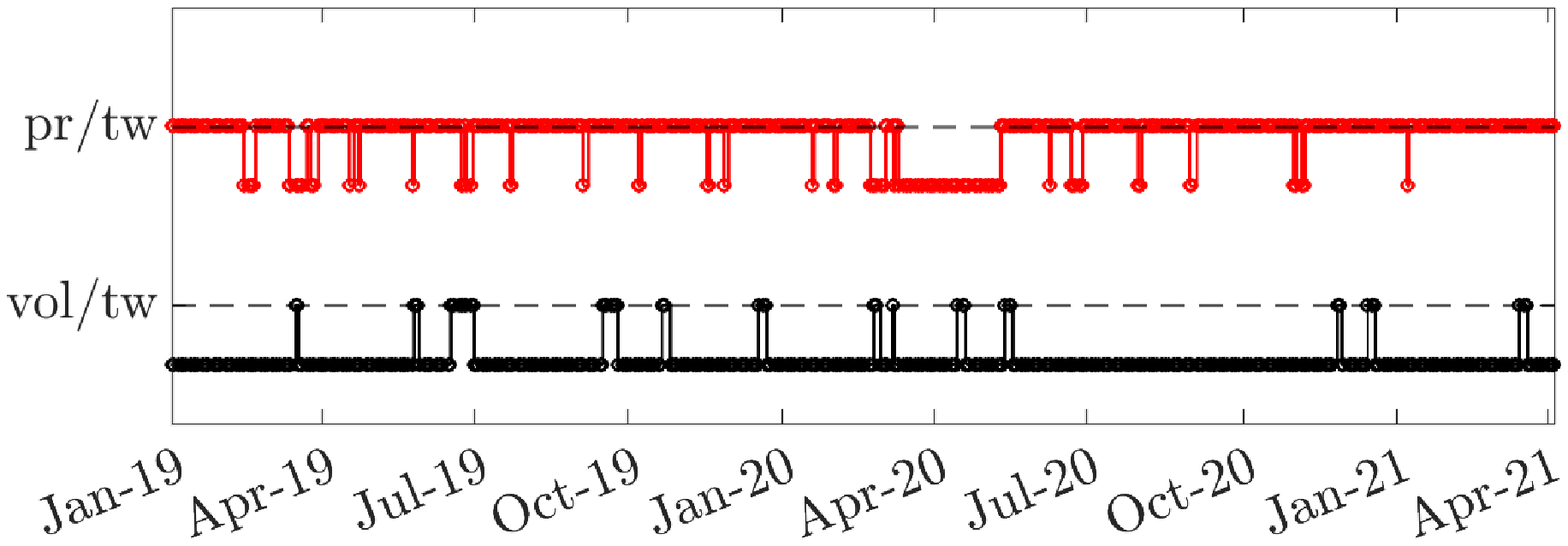} &
\vspace{-10ex}
\begin{minipage}[c]{0.3\textwidth}
\footnotesize
no period detected
\end{minipage} \vspace{10ex} \\[-5pt]
\begin{rotate}{90} \hspace{36pt} {\footnotesize PFE} \end{rotate} \hspace{-7pt} &
\includegraphics[trim= 5mm 0mm 1mm 0mm,clip,height= 3.1cm, width= 10.2cm]{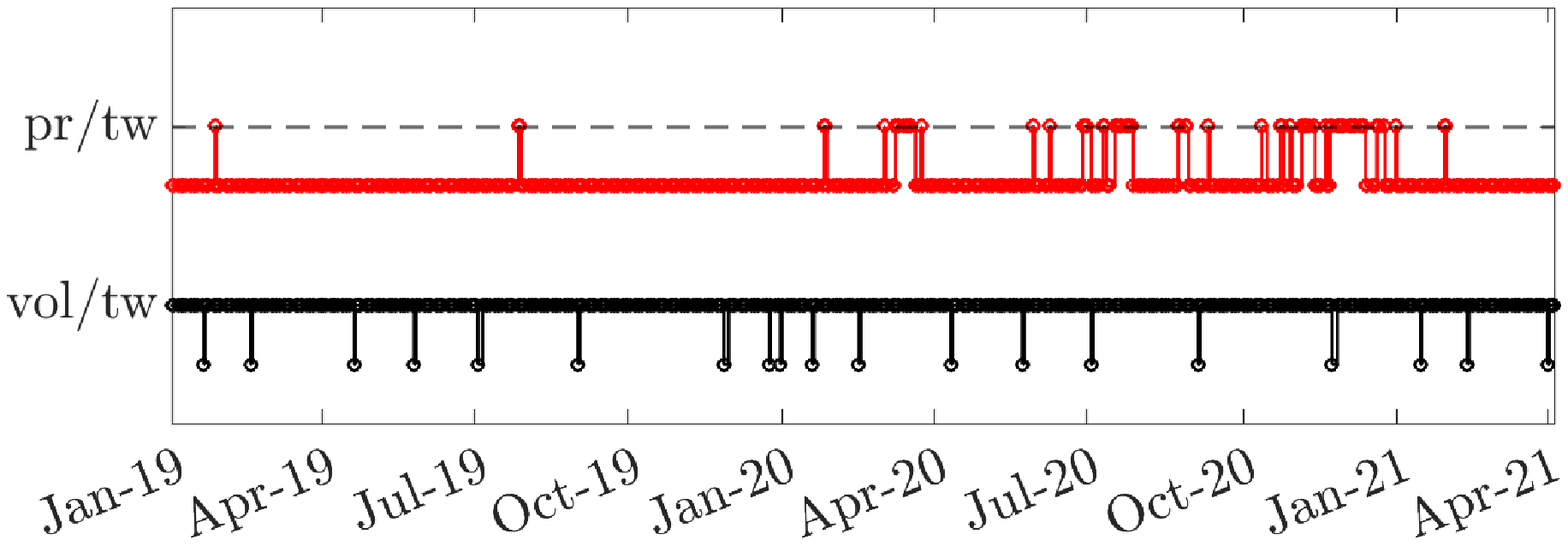} &
\vspace{-10ex}
\begin{minipage}[c]{0.3\textwidth}
\footnotesize
no period detected
\end{minipage} \vspace{10ex} \\[-5pt]
\begin{rotate}{90} \hspace{36pt} {\footnotesize DIS} \end{rotate} \hspace{-7pt} &
\includegraphics[trim= 5mm 0mm 1mm 0mm,clip,height= 3.1cm, width= 10.2cm]{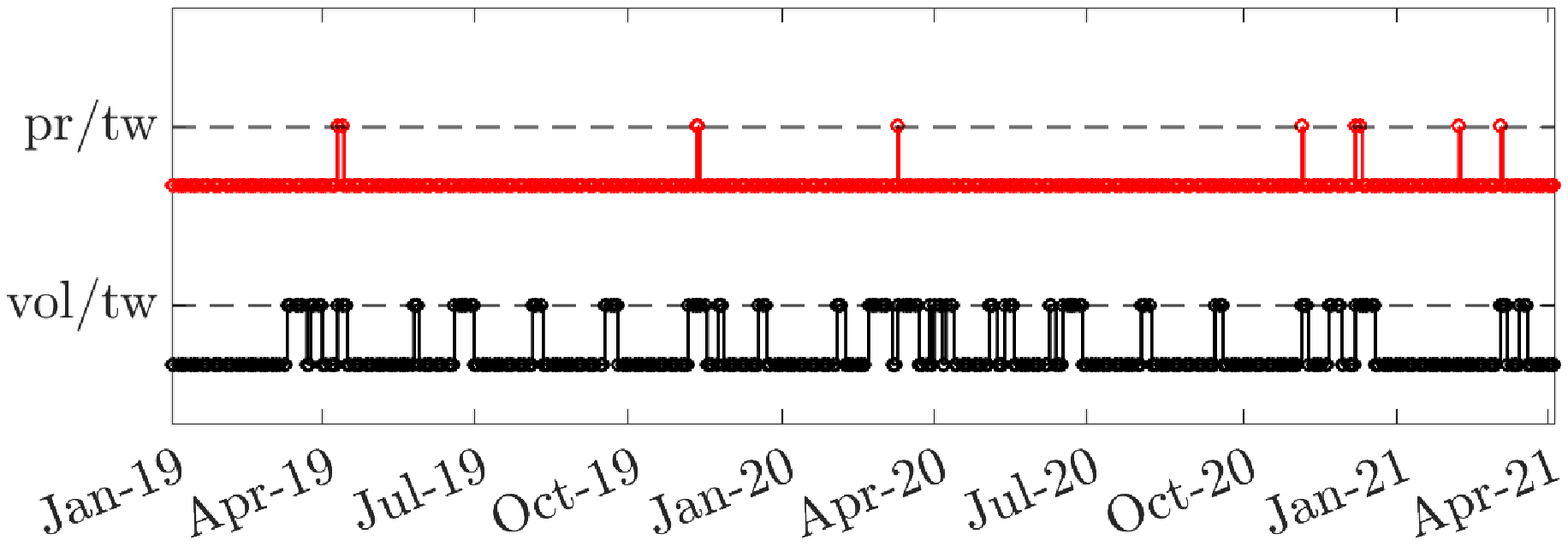} &
\vspace{-10ex}
\begin{minipage}[c]{0.3\textwidth}
\footnotesize
no period detected
\end{minipage} \vspace{10ex}
\end{tabular}
\caption{Estimated time-varying regimes, $\hat{S}_t$, for the price-tweets (pr/tw, red) and volumes-tweets (vol/tw, black) models. The dashed line indicates the cointegration regime, the gray shades represent the meme periods. Stocks: GameStop (GME), AMC Entertainment (AMC), KOSS Corporation (KOSS), Moody's (MCO), Pfizer (PFE), and Disney (DIS).}
\label{fig:plot_time_series}
\end{figure}
\FloatBarrier

In our empirical analysis, we impose minimum requirements of persistence and delay by fixing $d_c = 2$, $d_p = 2$, $d_c = 1$, and $d_f = 1$. The findings are robust to alternative choices of these parameters.

Figure \ref{fig:plot_time_series} shows for each stock the dynamics of the inferred cointegration relationships, as well as the mementum identified using the conditions in Definition \ref{def:meme_period}.
Each of the three selected stocks that were labeled by journalists as meme stocks (GME, AMC, KOSS) have experienced at least a meme period. The first stock to exhibit a mementum in 2021 is GameStop, on January 13, followed by AMC, on January 15, and KOSS, on January 25.\footnote{For example, the first day for GameStop's mementum, the stock's price exhibited a 57.39\% increase and traded volumes jumped from 7,060,665 to 144,501,736.}

\begin{figure}[t!h]
\centering
\includegraphics[trim= 0mm 0mm 0mm 0mm,clip,height= 5.5cm]{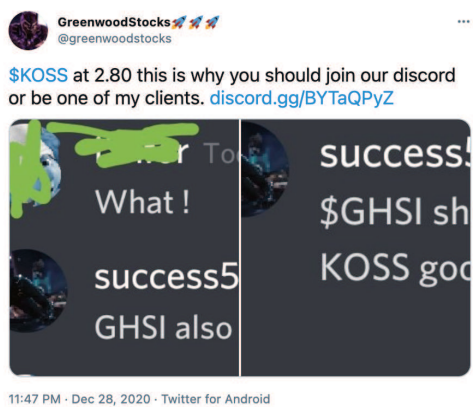} 
\caption{
Example of ``meme Tweet'' mentioning KOSS posted at the end of 2020. Tweet link: \url{https://twitter.com/greenwoodstocks/status/1343690011678486529}}
\label{fig:tweet_meme2}
\end{figure}

Interestingly, a 4-day meme period is identified for KOSS at the end of 2020. At that time the concept of meme stock had just been coined by journalists, and market commentators were starting questioning which mechanisms, unrelated to fundamentals, could have been at play behind the sharp rise of KOSS' stock price and trading volumes.\footnote{See  \href{https://marketglobalist.com/2020/12/29/is-there-any-reason-behind-the-dramatic-surge-of-koss-corporation-koss-stock/}{https://marketglobalist.com/2020/12/29/is-there-any-reason-behind-the-dramatic-surge-of-koss-corporation-koss-stock/}.} Moreover, during those days some ``meme tweets'' instrumental to investor coordination were already observable (see Figure \ref{fig:tweet_meme2}).

On the other hand, Moody's, Pfizer, and Disney, despite exhibiting multiple transitions between cointegrated and non-cointegrated regimes, don't have any meme period due to lack of synchronization and persistence. This confirms that the proposed framework is able to discriminate a mementum from other types of events that affect prices and/or traded volumes.
Differently from meme periods, these events may become viral but are not endogenous to social media.

\section{Conclusions}
The meme stock phenomenon has exerted a significant impact on financial markets, raising new challenges and policy issues that need to be addressed \cite{warren2021}.

Our contribution represents an initial attempt in this direction, by providing an econometric characterization of meme periods based on regime-switching cointegration and a simple procedure to identify them from market and social data.
We show that meme stocks are characterized by transitory periods where the coordination mechanism originated in social media is cointegrated with the stock's price and volumes.

The potentially destabilizing effect of social investors as an organized entity requires further investigation, also in terms of asset pricing and market efficiency.

\bibliographystyle{plain}
\bibliography{biblio}

%
%
%
%
%
%

\end{document}